\documentclass[12pt, prd, showpacs]{revtex4}
\usepackage{amssymb}
\usepackage{amsmath}

\input{tcilatex}

\begin{document}

\title{Equatorial geodesics in ergoregion of dirty black holes and zero
energy observers}
\author{O. B. Zaslavskii}
\affiliation{Department of Physics and Technology, Kharkov V.N. Karazin National
University, 4 Svoboda Square, Kharkov 61022, Ukraine}
\affiliation{Institute of Mathematics and Mechanics, Kazan Federal University, 18
Kremlyovskaya St., Kazan 420008, Russia}
\email{zaslav@ukr.net }

\begin{abstract}
We consider equatorial motion of particles in the ergoregion of generic
axially symmetric rotating black holes. We introduce the notion of zero
energy observers (ZEOs) as counterparts to known zero angular observers
(ZAMOs). It is shown that the trajectory of a ZEO has precisely one turning
point that lies on the boundary of the ergoregion for photons and inside the
ergoregion for massive particles. As a consequence, such trajectories enter
the ergosphere from the white hole region under horizon and leave it
crossing the horizon again (entering the black hole region). The angular
velocity of ZEO does not depend on the angular momentum. For particles with $%
E>0$ this velocity is bigger than for a ZEO, for $E<0$ it is smaller.
General limitations on the angular momentum are found depending on whether
the trajectory lies entirely inside the ergoregion, bounces back from the
boundary or intersects it. These results generalize the recent observations
made in A. A. Grib, Yu. V. Pavlov, arXiv:1601.02592 for the Kerr metric. We
also show that collision between a ZEO and a particle with $E\neq 0$ near a
black hole can lead to the unbound energy in the centre of mass thus giving
a special version of the Ba\~{n}ados - Silk - West effect.
\end{abstract}

\maketitle

\section{Introduction}

Motion of particle in the ergoregion possesses some peculiarities that are
absent in the outer region. Especially interesting is that the negative and
zero values of the Killing energy $E$ become possible. Some general
properties of trajectories with $E<0$ were considered in \cite{g3}, \cite%
{gpneg} for the Kerr metric and in \cite{neg} in much more general
background. Recently, the results of \cite{g3} for the Kerr metric were
extended \cite{E} to include the case $E=0$. In the present paper, we
generalize the analysis of \cite{E} to generic axially symmetric rotating
black holes. Real astrophysical black holes are surrounded by matter (they
are "dirty" in this sense) or electromagnetic fields, so their metric can
deviate from the Kerr form. Even if in practical astrophysics environmental
effects are small \cite{cp}, the notion of dirty black holes represents
conceptual issue important for thermodynamics including the problem of black
hole entropy, properties of the event horizons, hairy black holes, etc \cite%
{vis1} - \cite{bz}.

We derive some general inequalities relating the energy and angular momentum
depending on whether a trajectory intersects the boundary of the ergoregion,
lies entirely inside or bounces from the boundary back into the inner
region. We make main emphasis on trajectories with $E=0$ and argue that they
can be considered as counterparts of well known zero angular momentum
observers (ZAMOs) \cite{72} adapted to motion in the ergoregion.

Apart from general properties of individual trajectories, we consider
collisions of two particles near a black hole one of which has $E\approx 0$.
It turns out that this can give rise to the unbounded energy in the centre
of mass frame. The corresponding scenario was absent from \cite{ban} and
extends the list of possibilities of getting high $E_{c.m.}$

Below, we put fundamental constants $G=c=1$.

\section{Metric and equations of motion}

Let us consider the stationary axially symmetric metric

\begin{equation}
ds^{2}=-N^{2}dt^{2}+R^{2}(d\phi -\omega dt)^{2}+\frac{dr^{2}}{A}+g_{\theta
\theta }d\theta ^{2},  \label{met}
\end{equation}%
where all coefficients do not depend on $t$ and $\phi $. The form (\ref{met}%
) implies that we consider only spacetimes invariant to the simultaneous
inversion of the time and the azimuthal angle (see Ch. 2, Sec. 11 of \cite%
{ch} for details). We are interested in equations of motion for test
geodesic particles. In the Kerr metric, the variables can be separated \cite%
{car} and this simplifies analysis greatly. In a general case, the
separation of variables is impossible. To avoid complication, we assume that
the spacetime is symmetric with respect to the equatorial plane and we
restrict ourselves by consideration of motion in the equatorial plane $%
\theta =\frac{\pi }{2}$ only. Then, the equations of geodesics give us%
\begin{equation}
m\dot{t}=\frac{X}{N^{2}}\text{,}  \label{t}
\end{equation}%
\begin{equation}
m\dot{\phi}=\frac{L}{R^{2}}+\frac{\omega X}{N^{2}}\text{,}  \label{fi}
\end{equation}%
where dot denotes derivative with respect to the proper time $\tau $. Here, $%
m$ is the mass,%
\begin{equation}
X=E-\omega L\text{,}  \label{x}
\end{equation}%
$E=-mu_{0}$ is the energy, $L=mu_{\phi }$ being the angular momentum, $%
u^{\mu }=\frac{dx^{\mu }}{d\tau }$ the four-velocity. One can also use the
canonical parameter $\lambda $ along geodesics according to $\lambda =\tau
/m $ that is convenient for consideration of a massless limit.

The forward-in-time condition $\dot{t}>0$ implies that 
\begin{equation}
X\geq 0\text{.}  \label{x+}
\end{equation}%
According to (\ref{t}), the equality is possible on the horizon, where $N=0$%
. Outside it, $X>0$. Condition (\ref{x+}) has important physical
consequences in our context. In particular, for the value $E=0$ which we are
interested in (see below) it restricts the allowed range of $L$ forbidding
positive $L$.

The quantities $E$ and $L$ are conserved due to independence of the metric
of $t$ and $\phi $, respectively. From the normalization condition $u_{\mu
}u^{\mu }=-1$, one obtains that%
\begin{equation}
m\dot{r}=\pm \frac{\sqrt{A}}{N}Z\text{,}  \label{r}
\end{equation}%
where

\begin{equation}
Z^{2}=X^{2}-N^{2}(m^{2}+\frac{L^{2}}{R^{2}})\text{.}  \label{z}
\end{equation}%
It follows from 
\begin{equation}
Z^{2}\geq 0  \label{z+}
\end{equation}%
and (\ref{x}) that%
\begin{equation}
E\geq \omega L+N\sqrt{m^{2}+\frac{L^{2}}{R^{2}}}\text{.}  \label{el}
\end{equation}

It is worth noting that in passing from (\ref{z+}) to (\ref{el}), we
rejected the inequality with the opposite sign \ since it would violate
condition (\ref{x+}).

Within the plane $\theta =\frac{\pi }{2}$ under consideration, we can always
redefine the radial coordinate in such a way that $A=N^{2}.$ Then, we have%
\begin{equation}
m\dot{r}=\sigma Z\text{,}  \label{rad}
\end{equation}%
$\sigma =\pm 1$.

Also, it follows from (\ref{t}), (\ref{fi}) and (\ref{rad}) that%
\begin{equation}
\frac{d\phi }{dt}=\omega +\frac{LN^{2}}{R^{2}X}\text{,}  \label{ft}
\end{equation}%
\begin{equation}
\frac{dr}{dt}=\frac{\sigma ZN^{2}}{X}\text{.}  \label{rt}
\end{equation}

In the metric (\ref{met}) we imply that $\omega $ changes sign nowhere. In
particular, this happens for the Kerr and Kerr-Newman metrics. Then, one can
always achieve $\omega >0$ (if $\omega <0$, it is sufficient to make
substitution $\phi \rightarrow -\phi $). The situation with $\omega $
changing sign is not forbidden, in principle, but we do not discuss such
more involved situations.

\section{Ergoregion, energy and angular momentum}

For the metric (\ref{met}),%
\begin{equation}
g_{00}=-N^{2}+R^{2}\omega ^{2}\text{.}  \label{00}
\end{equation}

By definition, the boundary of the ergoregion is located where%
\begin{equation}
g_{00}=0,\text{ }N^{2}=R^{2}\omega ^{2}\text{.}  \label{b}
\end{equation}

Inside the ergosphere, 
\begin{equation}
g_{00}>0\text{, }N^{2}<R^{2}\omega ^{2}\text{.}  \label{in}
\end{equation}%
Outside it, 
\begin{equation}
g_{00}<0,\text{ }N^{2}>R^{2}\omega ^{2}.  \label{out}
\end{equation}%
The properties of a trajectory depend strongly on where it is located.

It follows from (\ref{el}), (\ref{out}) that outside the ergoregion $E>0$.
Inside the ergoregion all cases $E>0$, $E=0$ and $E<0$ are possible. General
properties of geodesics with $E<0$ are described in \cite{g3} for the Kerr
metric and, for equatorial motion, are generalized in \cite{neg}. Now, we
make main emphasis on the trajectories with $E=0$.

\section{Properties of trajectories with $E=0$}

Now, we prove the following statements concerning trajectories with $E=0$
that generalize those of \cite{E}.

1) The turning point lies on the boundary of the ergoregion for massless
particles and inside it for massive ones.

2) There is exactly one turning point.

3) There are no circular orbits.

By substitution into (\ref{z}), we have%
\begin{equation}
Z^{2}=L^{2}(\omega ^{2}-\frac{N^{2}}{R^{2}})-m^{2}N^{2}=\frac{L^{2}g_{00}}{%
R^{2}}-m^{2}N^{2}\text{.}  \label{z0}
\end{equation}

In the turning point, $\dot{r}=0$, so $Z=0$ according to (\ref{rad}). If $%
m=0 $, we see that $Z>0$ everywhere inside the ergosphere due to (\ref{in}),
so there are no turning points there. Only on the boundary (\ref{b}) $Z=0$.
If $m>0$, there is a turning point inside where $\frac{L^{2}g_{00}}{R^{2}}%
=m^{2}N^{2}$. Thus statement 1) is proved.

Let us assume that%
\begin{equation}
N^{\prime }>0\text{, }\omega ^{\prime }<0\text{, }R^{\prime }>0\text{,}
\label{as}
\end{equation}%
\begin{equation}
(\omega R)^{\prime }<0.  \label{as1}
\end{equation}

Here, prime denotes derivative with respect to $r$. One can check that all
the assumptions (\ref{as}), (\ref{as1}) are valid for the Kerr and
Kerr-Newman metrics. They do not have a universal meaning but look quite
natural physically since they state that the metric approaches its flat
spacetime limit at infinity rapidly enough and in a quite "natural" way.
Namely, the assumptions (\ref{as}) state that the lapse function, areal
radius and rotation of spacetime monotonically change everywhere from the
horizon to infinity. Condition (\ref{as1}) is more strong assumption
according to which fall-off of rotation dominates over growth of the areal
radius, so the metric approaches its static limit more closely even before
it approaches the Minkowski form.

Now, we may take advantage of the approach of \cite{neg} where it was used
for negative energies. For $E=0$ it somewhat simplifies.

Taking the derivative we obtain%
\begin{equation}
\left( Z^{2}\right) ^{\prime }=-m^{2}(N^{2})^{\prime }+L^{2}(\omega
^{2})^{\prime }-L^{2}(\frac{N^{2}}{R^{2}})^{\prime }\text{.}
\end{equation}%
It follows from (\ref{b}), (\ref{in}) that%
\begin{equation}
-(\frac{N^{2}}{R^{2}})^{\prime }\leq -\frac{2NN^{\prime }}{R^{2}}+\frac{%
2\omega ^{2}R^{\prime }}{R}\text{,}
\end{equation}%
whence%
\begin{equation}
\left( Z^{2}\right) ^{\prime }\leq -m^{2}(N^{2})^{\prime }+2\frac{%
L^{2}\omega (\omega R)^{\prime }}{R}-\frac{2NN^{\prime }}{R^{2}}L^{2}<0
\label{impos}
\end{equation}%
due to (\ref{as}).

Thus $Z^{2}$ is a monotonically decreasing function of $r$. Therefore,
equation $Z=0$ has precisely one root, so there is only one turning point.
As is established above, it is located inside the ergoregion (for massive
particles) or on its boundary (for massless ones). Thus statement 2) is also
proved.

The existence of circular orbits requires $Z=0$ and $\left( Z^{2}\right)
^{\prime }=0$. However, this is also impossible because of (\ref{impos}).
Thus circular orbits are absent, so statement 3) is proved as well.

Properties 2 and 3 have an important consequence if we consider the complete
behavior of geodesics. As there are no circular orbits, and oscillating
trajectories between two turning points are also excluded, geodesics with $%
E=0$ (similarly to the case $E<0$) enter the ergosphere from the inner
region under the horizon radius (i.e. from the white hole region) and leave
it again entering the region under the horizon (into the black hole region).
This generalizes the corresponding result for the Kerr metric \cite{E}. See
also \cite{g3} and \cite{bw}.

It is also worth noting that (\ref{x+}) entails that for $E=0$ the angular
momentum $L<0$.

\section{General inequalities on E and L}

The expression (\ref{z}) can be rewritten as%
\begin{equation}
Z^{2}=\frac{g_{00}}{R^{2}}(L-L_{+})(L-L_{-})\text{,}  \label{zr}
\end{equation}%
where%
\begin{equation}
L_{\pm }=\frac{\omega ER^{2}}{g_{00}}\pm \frac{NR}{g_{00}}\sqrt{%
E^{2}+m^{2}g_{00}}=\frac{\omega ER^{2}}{g_{00}}\pm \frac{R\sqrt{\omega
^{2}R^{2}-g_{00}}}{g_{00}}\sqrt{E^{2}+m^{2}g_{00}}\text{.}  \label{l+}
\end{equation}

\subsection{Outside the ergoregion}

Now, $g_{00}<0$, it follows from (\ref{el}) and (\ref{out}) that $E>0$, as
it should be$.$ It is seen from (\ref{zr}) and (\ref{l+}) that now%
\begin{equation}
L_{+}\leq L\leq L_{-}.  \label{+-}
\end{equation}

It is also necessary that the expression inside the square root be
nonnegative. Otherwise, $Z^{2}$ would be negative. Thus we obtain%
\begin{equation}
E\geq m\sqrt{-g_{00}}\text{.}  \label{emg}
\end{equation}
This is in agreement with eq. (88.9) of \cite{ld}. It is easy to check that $%
L_{-}<\frac{E}{\omega }$, so if (\ref{+-}) is satisfied, eq. (\ref{x+}) is
satisfied as well.

\subsection{Inside the ergoregion}

Now, $g_{00}>0$. Then, (\ref{z+}) entails $L\geq L_{+}$ or $L\leq L_{-}$.
However, condition (\ref{x+}) excludes the first variant since it is easy to
check that now $L_{+}\geq \frac{E}{\omega }$. Thus the only possible variant
is%
\begin{equation}
L\leq L_{-}=\frac{\omega ER^{2}}{g_{00}}-\frac{NR}{g_{00}}\sqrt{%
E^{2}+m^{2}g_{00}}\text{,}  \label{min}
\end{equation}%
where equality is achieved at the turning point.

Now, all cases $E<0$, $E=0$ and $E>0$ are possible. In particular, for $E=0$%
, we obtain from (\ref{min}) that%
\begin{equation}
L\leq -\frac{NRm}{\sqrt{g_{00}}}\text{.}  \label{2}
\end{equation}

\subsection{On the boundary}

Using (\ref{b}) on the boundary, we obtain from (\ref{el})

\begin{equation}
E\geq \omega _{0}(L+\sqrt{m^{2}R_{0}^{2}+L^{2}})\geq 0\text{.}  \label{pos}
\end{equation}%
Let us consider the limit $g_{00}\rightarrow 0$ for $L_{\pm }$ taken, say,
in the outer region. It is seen from (\ref{l+}) that in this case 
\begin{equation}
L_{+}\rightarrow -\infty \text{, }L_{-}\rightarrow \frac{E}{2\omega _{0}}-%
\frac{m^{2}\omega _{0}R_{0}^{2}}{2E}\text{,}
\end{equation}%
where subscript "0" refers to the boundary $r=r_{0}$ of the ergoregion.

Thus 
\begin{equation}
L\leq \frac{E}{2\omega _{0}}-\frac{m^{2}\omega _{0}R_{0}^{2}}{2E}\text{.}
\label{ineq}
\end{equation}

One can also consider the same limit taken from inside and obtain the same
result (\ref{ineq}). It can be obtained from (\ref{z}) directly with (\ref{b}%
) taken into account. Then,%
\begin{equation}
Z^{2}=E^{2}-2\omega EL+\frac{L^{2}g_{00}}{R^{2}}-m^{2}N^{2}
\end{equation}%
and (\ref{z+}) gives us (\ref{ineq}).

Impossibility of $E<0$ that follows from (\ref{pos}) can be noticed also
from (\ref{min}) since $E<0$ for a trajectory approaching the boundary would
lead to $\lim_{g_{00}\rightarrow +0}L_{-}=-\infty $ thus violating (\ref{min}%
).

The trajectories with $E>0$ are possible for massive and massless particles.
The case $E=0$ can be realized for $m=0$ only (for the Kerr metric, this was
noticed in \cite{E}). Then, $L<0$ is arbitrary. Formally, (\ref{ineq}) for
massless particles admits also $E=0$, $L=0$ but this is inconsistent with (%
\ref{x+}) outside the horizon although on the horizon itself such a
trajectory is possible \cite{circD}.

\section{Angular velocity}

Now, we will find some general properties of geodesics depending on the
energy.

\subsection{$E=0$}

The angular velocity of any particle with $E=0$ (massive or massless) does
not depend on the angular momentum. This is valid inside the ergoregion and
on its boundary, where it vanishes.

\textit{Proof. }By substitution $E=0$ into (\ref{x}) and (\ref{ft}), we
obtain%
\begin{equation}
\left( \frac{d\phi }{dt}\right) _{E=0}=\frac{g_{00}}{\omega R^{2}}\text{,}
\label{f0}
\end{equation}%
where (\ref{00}) was taken into account. We see that indeed $L$ drops out
from the formula. Everywhere inside the ergoregion, $\left( \frac{d\phi }{dt}%
\right) _{E=0}>0$ as it should be. On the boundary, 
\begin{equation}
\left( \frac{d\phi }{dt}\right) _{E=0}=0
\end{equation}%
according to (\ref{b}).

\subsection{$E\neq 0$}

The sign of the difference $\left( \frac{d\phi }{dt}\right) _{E}-(\frac{%
d\phi }{dt})_{0}$ is defined by that of $E$.

\textit{Proof.} Subtracting (\ref{f0}) from (\ref{ft}), we obtain%
\begin{equation}
\left( \frac{d\phi }{dt}\right) _{E}-(\frac{d\phi }{dt})_{E=0}=\frac{N^{2}E}{%
R^{2}\omega X}\text{.}
\end{equation}%
For $E<0$ the angular velocity is smaller than for $E=0$ and for $E>0$ it is
bigger. This generalizes observation made in \cite{E} for the Kerr black
hole.

\subsection{Zero energy versus zero angular momentum observers}

Bearing in mind all properties described above, it is natural to introduce
the zero-energy observer (ZEO) by analogy with zero angular momentum (ZAMO)
ones \cite{72}. For a ZAMO, $\frac{d\phi }{dt}=\omega $ is determined by the
metric entirely and does not depend on the energy. Then, the difference $%
\frac{d\phi }{dt}-\left( \frac{d\phi }{dt}\right) _{ZAMO}$ is determined by
the sign of $L$ completely according to (\ref{ft}) since always $X>0$ (\ref%
{x+}). This is direct counterpart of the corresponding properties of ZEO
discussed in this Section, with $E$ replaced with $L$.

Meanwhile, there is some difference between two groups of these observers.
ZAMOs are not geodesics since they imply $r=const$, so (\ref{r}) is,
generally speaking, not satisfied. By contrary, the trajectories with $E=0$
under discussion are geodesics. Moreover, circular orbits $r=const$ are
impossible for them. In this respect, ZEO and ZAMO can be considered as
complimentary to each other. From the other hand, ZEOs are possible in the
ergoregion only whereas there is no such a restriction on ZAMOs.

\section{Comparison to the Kerr metric}

For the Kerr metric in the Boyer-Lindquiste coordinates in the plane $\theta
=\frac{\pi }{2}$ one has%
\begin{equation}
R^{2}=r^{2}+a^{2}+\frac{2M}{r}a^{2}\text{,}
\end{equation}

\begin{equation}
\omega =\frac{2Ma}{R^{2}r},
\end{equation}%
\begin{equation}
N^{2}=\frac{\Delta }{R^{2}}\text{, }g_{00}=-(1-\frac{2M}{r})\text{,}
\label{nr}
\end{equation}%
\begin{equation}
\Delta =r^{2}-2Mr+a^{2}.
\end{equation}%
Taking these quantities on the boundary of the ergoregion $r=2M$, one
obtains from (\ref{ineq}) that%
\begin{equation}
L\leq E(a+\frac{2M^{2}}{a})-\frac{m^{2}a}{2E}\text{.}
\end{equation}%
This coincides with eq. (22) of \cite{E} if one puts there $\theta =\frac{%
\pi }{2}$ and $Q=0$ (where $Q$ is the Carter constant).

Eq. (\ref{f0}) turns into 
\begin{equation}
\left( \frac{d\phi }{dt}\right) _{E=0}=\frac{r-2M}{2Ma}
\end{equation}%
that coincides with eq. (35) of \cite{E}.

Eq. (\ref{2}) gives us%
\begin{equation}
L\leq -\frac{m\sqrt{\Delta }}{\sqrt{\frac{2M}{r}-1}}\text{,}  \label{minkerr}
\end{equation}%
where we took into account (\ref{nr}). Eq. (\ref{minkerr}) coincides with
eq. (33) of \cite{E} for $\theta =\frac{\pi }{2}$ and $Q=0$.

\section{Properties of photon orbits}

Here, we describe some properties of trajectories of massless particles
(photons). It follows from (\ref{ft}), (\ref{rt}) that%
\begin{equation}
\frac{d\phi }{dr}=\sigma \frac{E\omega R^{2}-Lg_{00}}{ZN^{2}R^{2}}\text{.}
\end{equation}

This expression admits the zero mass limit $m\rightarrow 0$. Let also $E=0$.
Then, $L=-\left\vert L\right\vert $, $X=\omega \left\vert L\right\vert $, it
is seen from (\ref{z0}) that 
\begin{equation}
Z=\frac{\left\vert L\right\vert }{R}\sqrt{g_{00}}\text{.}
\end{equation}%
Then, we obtain%
\begin{equation}
\frac{d\phi }{dr}=\sigma \frac{\sqrt{g_{00}}}{RN^{2}}\text{,}
\end{equation}%
so the angular momentum drops out from the formula similarly to the Kerr
case \cite{E} .

Now, the upper point of the trajectory with $E=0$ is on the boundary of the
ergoregion $r=r_{0}$. Near it, $g_{00}\approx B(r_{0}-r)$, where $B>0$ is
some constant, so%
\begin{equation}
\phi \approx \frac{2}{3}\frac{\sigma }{R(r_{0})N^{2}(r_{0})}\sqrt{B}%
(r_{0}-r)^{3/2}+const\text{.}
\end{equation}%
Thus there is a point of inflexion. One can check that this is in agreement
with the properties of eq. (39)\ of \cite{E}.

\section{Collisions}

It turns out that consideration of particles with zero (or extremely small)
energy suggests a new option in the theory of high energy collisions that
remained unnoticed before. Let two particles 1 and 2 collide. Then, in the
point of collision, one can define the energy in the centre of mass frame
according to%
\begin{equation}
E_{c.m.}^{2}=-(m_{1}u_{1\mu }+m_{2}u_{2\mu })(m_{1}u_{1}^{\mu
}+m_{2}u_{2}^{\mu })=m_{1}^{2}+m_{2}^{2}+2m_{1}m_{2}\gamma \text{,}
\end{equation}%
where 
\begin{equation}
\gamma =-u_{1\mu }u^{2\mu }  \label{gamma}
\end{equation}%
is the Lorentz factor of their relative motion. Using the geodesic equations
(\ref{t}) - (\ref{z}) and substituting them directly into (\ref{gamma}), one
can obtain after summation over indices that

\begin{equation}
m_{1}m_{2}\gamma =\frac{X_{1}X_{2}-Z_{1}Z_{2}}{N^{2}}-\frac{L_{1}L_{2}}{R^{2}%
}\text{.}  \label{ga}
\end{equation}

Under certain conditions, $\gamma $ becomes unbounded. This is the Ba\~{n}%
ados, Silk and West (BSW) effect \cite{ban}. This happens, if collisions
occurs near the horizon ($N=0$) and (i) one of particles has such fine-tuned
parameters that $X_{H}=0$, (ii) the second particle is not fine-tuned, $%
X_{H}>0$ for it (this is so-called usual particle) \cite{prd}.\ Here,
subscript "H" means that the corresponding quantity is calculated on the
horizon. Now we will see that inclusion of trajectories with $E=0$ into
consideration extends this scheme.

Let $E_{1}=0$ and $L_{1}$ be extremely small and negative. Namely, we assume
that 
\begin{equation}
\left\vert L_{1}\right\vert \omega _{H}=\alpha m_{1}N_{c}\text{,}  \label{al}
\end{equation}%
where $\alpha =O(1)$, $\alpha >1$, subscript "c" refers to the point of
collision near the horizon, $N_{c}\ll 1$.

Then, (\ref{ga}) \ gives us that%
\begin{equation}
m_{1}m_{2}\gamma \approx \frac{X_{2}m_{1}}{N_{c}}(\alpha -\sqrt{\alpha ^{2}-1%
})
\end{equation}%
becomes unbounded, $\frac{E_{cm.}^{2}}{m_{1}^{2}}=O(\frac{X_{2}}{m_{1}N_{c}}%
)\gg 1$, provided $m_{1}N_{c}\ll X_{2}$. The same is true if $E_{1}\neq 0$
but $\left\vert E_{1}\right\vert \ll \left\vert L_{1}\right\vert \omega _{H}$%
. Condition (\ref{z+}) is satisfied in a narrow but nonzero strip where $%
0<N\lesssim \alpha N_{c}$.

It is instructive to compare the situation to that with participation of
near-critical particles in the standard BSW scenario \cite{ban}, \cite{prd}.
In the latter case, let particle 1 be "near-critical" in the sense that $%
E_{1}=\omega _{H}L_{1}+O(N_{c})$. Then, taking into account that $\omega
-\omega _{H}=O(N_{c})$ for extremal black holes \cite{tan}, we can see that $%
X_{1}=O(N_{c})$. One obtains from (\ref{ga}) that, if particle 2 is usual, $%
\gamma =O(N_{c}^{-1})$ is unbounded. In doing so, in the limit $%
N_{c}\rightarrow 0$, the energy $E_{1}\rightarrow \omega _{H}L_{1}$, $\left(
X_{1}\right) _{H}\rightarrow 0$, so near-critical particle becomes exactly
critical. However, now such near-critical trajectories do not have the
critical ones as their limit when $N_{c}\rightarrow 0$. Indeed, $\left\vert
L_{1}\right\vert \omega _{H}\gg \left\vert E_{1}\right\vert $, so $E_{1}\neq
\omega _{H}L_{1}$ and we see that $X_{H}\neq 0$. Thus both particles 1 and 2
are usual. In the scenario under discussion, a black hole can be extremal or
nonextremal, so this does not require multiple collisions in contrast to the
standard scenarios near nonextremal black holes \cite{gp}.

\subsection{Kinematic explanation}

There is a relation \cite{k}%
\begin{equation}
X=E-\omega L=\frac{mN}{\sqrt{1-V^{2}}}\text{,}
\end{equation}%
where $V$ is the velocity measured by a ZAMO. When a point of collision
approaches the horizon, $N\rightarrow 0$ and, in a general case, $%
V\rightarrow 1$. However, for a fine-tuned (critical) particle with $%
E=\omega _{H}L+O(N)$, we have $V<1$. Thus we have collision between a rapid
and slow particles that leads to the relative velocity $w\rightarrow 1$, so $%
\gamma $ diverges (see \cite{k} for details). Now, $X$ is also of the order $%
N$ but not due to special relation between $E$ and $L$ but due to (\ref{al})
together with $E=0$, hence again $V<1$ and $w\rightarrow 1$.

\section{Conclusion}

Thus we derived some general restriction on the relation between the angular
momentum and energy depending on whether or not a geodesics intersects the
boundary of the ergoregion. The results apply to any axially symmetric
rotating metrics. In the particular case of the Kerr one the results of \cite%
{E} are reproduced. It turned out that trajectories with $E=0$ contain
precisely one turning point inside or on the boundary, the latter case being
possible for massless particles only. The full history of geodesics with $%
E=0 $ is such that they originate from the white hole region under the
horizon and return to the black hole region thus realizing behavior typical
of a black-white eternal hole.

We showed that zero energy observes (ZEO) can be considered as counterparts
to well known zero angular momentum observers (ZAMO) adapted just to motion
in the ergoregion. Their angular velocity does not depend on $L$ and in this
sense represents a natural reference point for comparison with angular
velocities of other particles. The sign of the difference of angular
velocities with $E\neq 0$ and $E=0$ is determined by $E$ entirely.

We also found a new scenario of high energy collisions. It represents
modification of the standard BSW effect with participation of ZEOs or
particles which differ from it slightly having nonzero but extremely small
energy.

It would be of interest to extend all or some of these generic results to
nonequatorial motion.

\begin{acknowledgments}
This work was funded by the subsidy allocated to Kazan Federal University
for the state assignment in the sphere of scientific activities.
\end{acknowledgments}

\end{document}